\documentclass[final,5p,times,twocolumn]{elsarticle}

\usepackage{amssymb}
\usepackage{amsmath,amsfonts}
\usepackage{algorithm}
\usepackage{algorithmicx}
\usepackage{algpseudocode}
\usepackage{array}
\usepackage[caption=false,font=normalsize,labelfont=sf,textfont=sf]{subfig}
\usepackage{textcomp}
\usepackage{stfloats}
\usepackage{url}
\usepackage{verbatim}
\usepackage{graphicx}
\usepackage{latexsym}
\usepackage{bm}
\usepackage{amsfonts}
\usepackage{amsmath}
\usepackage{multirow}
\usepackage{caption}
\usepackage{xcolor}
\usepackage{colortbl}
\usepackage{booktabs}
\usepackage{subfloat}
\biboptions{sort&compress}
\graphicspath{{./graphics/}}

\journal{Neurocomputing}

\begin{document}

\begin{frontmatter}


\title{Instance Attack:An Explanation-based Vulnerability Analysis Framework Against DNNs for Malware Detection}

\author[1]{Sun RuiJin} 
\author[4]{Guo ShiZe}
\author[5]{Guo JinHong}
\author[1]{Xing ChangYou}
\author[2]{Yang LuMing}
\author[3]{Guo Xi\corref{cor1}}
\author[1]{Pan ZhiSong\corref{cor1}}
\cortext[cor1]{Corresponding author}
\affiliation[1]{Army Engineering University of PLA}
\affiliation[2]{National University of Defense Technology}
\affiliation[3]{University Of Science & Technology Beijing}
\affiliation[4]{National Computer Network and Information Security Management Center}
\affiliation[5]{Shanghai Jiao Tong University}

\begin{abstract}
	Deep neural networks (DNNs) are increasingly being applied in malware detection and their robustness has been widely debated. Traditionally an adversarial example generation scheme relies on either detailed model information (gradient-based methods) or lots of samples to train a surrogate model, neither of which are available in most scenarios. 
	We propose the notion of the instance-based attack. Our scheme is interpretable and can work in a black-box environment. Given a specific binary example and a malware classifier, we use the data augmentation strategies to produce enough data from which we can train a simple interpretable model. We explain the detection model by displaying the weight of different parts of the specific binary. By analyzing the explanations, we found that the data subsections play an important role in Windows PE malware detection. We proposed a new function preserving transformation algorithm that can be applied to data subsections. By employing the binary-diversification techniques that we proposed, we eliminated the influence of the most weighted part to generate adversarial examples. Our algorithm can fool the DNNs in certain cases with a success rate of nearly 100\%. Our method outperforms the state-of-the-art method introduced in \cite{2019Optimization}. The most important aspect is that our method operates in black-box settings and the results can be validated with domain knowledge. Our analysis model can assist people in improving the robustness of malware detectors.
\end{abstract}

\end{frontmatter}
\section{INTRODUCTION}
Malware attacks is an important issue for the cybersecurity society nowadays. Thousands of malware attacks are reported every day, according to \citep{2020Fuctionality}'s description. 
Both academia and industry have devoted a lot of manpower to malware detection. Traditional detection methods, such as SVM \citep{2016Detecting} and signature \citep{2012MOMENTUM} require manual feature engineering, which is challenging. Due to the large amount of malware, the manual work is time consuming and tedious. As DNNs have made significant advances in many domains, including image \citep{2016Accessorize} and voice classification \citep{2019Imperceptible}, more and more researchers and anti-virus enterprises adopt DNN-based detectors in the field of cybersecurity. The DNNs models automatically make the classification for malware without expert knowledge. Researchers use deep learning models in an end-to-end manner that operates directly on the raw bytes of Windows Portable Executable (PE) files.\\
In the cybersecurity field, malware detection systems can be categorized into dynamic and static. Dynamic detection systems learn behavioral features that can be used for classification, while static ones classify files using features directly without execution \citep{2019Optimization}. We focus on the static methods in this paper. Several byte-based DNNs models have shown competitiveness with the traditional models \citep{2015Deep}\citep{article}. 
The robustness of the DNNs detection system and the interpretability of DNNs have attracted much attention, while the DNNs have shown great potential. Model interpretability is especially important in finance and security-related areas. Lack of interpretability limits their application.
Adversarial examples are the techniques that focuse on perturbing the examples to mislead DNNs detection systems. They help to improve the robustness of DNNs. 
Unlike other domains, the constraints of semantic invariance must be satisfied in binary. When we generate an adversarial example, we transform a malware and it would be classified as benign while we must keep its semantics unchanged. People introduce different transformation techniques that could keep the binaries' functionalities intact  \citep{DBLP:journals/corr/abs-1801-08917}\citep{DBLP:journals/corr/abs-2003-03100}\citep{DBLP:journals/corr/abs-1904-04802}. Transformations here mean the modificaions that can be made to a PE file that do not break the PE file format and do not alter code execution \citep{{DBLP:journals/corr/abs-1801-08917}}. Although researchers have made a lot of progress in generating adversarial examples for malware detection, there are still some unsolved problems. Firstly, only a few articles that use DNNs to detect malware have explained their detection results. The lack of transparency makes it questioned by many people \cite{DBLP:journals/corr/abs-2010-09470}. The unexplainable model may detect the binaries according to false causalities that are unrelated to any malicious activity \cite{DBLP:journals/corr/abs-2010-09470}. 
Secondly, the binary transformation methods that other people have adopted focus on the structural part\cite{2020Fuctionality} and the code part \cite{2019Optimization} but they leave the data section alone. Thirdly, after figuring out how to transform malware, they resort to complicated optimization methods (such as genetic algorithms) \cite{2020Fuctionality} or unexplainable stochastic methods \cite{2019Optimization}. These deficiencies limit their performance under the black-box model. Our approach fills the gap.\\
In this paper, we propose the notion of instance-based attacks. Our method is very similar to the transfer-based approach. The most important difference between instance-based and transfer-based methods is that all the data used to train our model is produced by data augmentation from one single binary.
We use an explanation based adversarial examples generation technology to test malware detectors by approximating the decision boundary iteratively. Our method is more effective compared with other methods in the context of black-box settings. In order to evade the DNNs in less steps, we could transform the most influential modules each round. In addition, our optimization method is explainable and can be verified with domain knowledge. We highlight our contributions below.
\begin{itemize}
	\item  We propose the idea of instance-based attack. By local reconstruction of decision boundaries, we do not train a surrogate model against the entire model, we train a substitute model for an instance. We focus on perturbing around a specific example to generate the adversarial examples by approximating the decision boundary iteratively.
	\item We introduce the interpretable model to analyze the results of typical detection models. In our model, we can quantify the weight of different parts for a specific binary in black-box settings. 
	\item We propose a novel functionality-preserving transformation method which is suitable for data sections in PE files that have not been evaluated by other authors.
	\item The theoretical and mathematical basis of our model are discussed in this paper. We explain in detail how we produce the augmentation data and how the model is trained.
	\item We tested our method in various scenarios and discovered that it outperformed other state-of-the-art methods in black-box settings\cite{2019Optimization}. Our method requires no human intervention. We can achieve a success rate of nearly 100\% in certain cases. Moreover we can verify the results with our domain knowledge.
\end{itemize}
\section{BACKGROUND AND RELATED WORK}
We start this section by providing an overview of popular DNNs-based malware detectors, we mainly focus on the static approaches. Then we discuss the existing evasion methods for malware detectors that target the raw bytes of PE files. We end the section by introducing articles on the explanation of DNNs for malware classification.
\subsection{DNNs for Malware Detection}
Malware detection plays an important role in the field of cyber security. DNNs have been used widely by researchers in malware classification. 
The most appealing aspect of the DNNs-based malware detectors is their ability to achieve state-of-the-art performance from raw bits rather than manually crafted features that require tedious human effort. Many DNNs-based detectors have been proposed so far, and we will introduce the most famous ones here. 
\citet{10.1145/2016904.2016908} visualize the malware binaries as gray-scale images. A classification method using standard image features is proposed with the observation that malware images belonging to the same family appear very similar in layout and texture. Then they use the classifier originally designed for images to sort malware. \citet{coull2019activation} introduce a DNN with five convolutional and pooling layers. It also has a learnable 10-dimensional embedding layer. At the end of the network, there is a single fully-connected layer and a sigmoid function.
\citet{2015Deep} employ four distinct complementary features from the static, benign and malicious binaries. They use a DNNs-based classifier which consists of an input layer, two hidden layers, and an output layer. They translate the outputs of the DNNs into a score that can be realistically interpreted as an approximation of the probability of whether the file is malware. \citet{articletRepresentation} proposes deep convolutional neural networks (CNN) that combine a ten-dimensional, learnable embedding layer with a series of five interleaved convolutional and max-pooling layers arranged hierarchically.
MalConv \citep{article} is the most popular CNN model which combines an eight-dimensional learnable embedding layer. \citet{article} have tried many different structures. They have attempted to use deeper networks (up to 13 layers), narrower convolutional filters (width 3–10), and smaller strides (1–10). 
Finally, they adopt the network that consists of two 1-D gated convolutional windows with 500 strides. We used the MalConv detection model to evaluate the effectiveness of our method.
\subsection{Adversarial Examples Against Malware Detector based on DNNs }
Adversarial examples are the technologies that focus on the minimum perturbations of input to break machine learning algorithms.
They can expose the vulnerability of the machine learning model and improve the robustness of the DNNs. For example, when DNNs are used in street sign detection, researchers show ways to mislead street signs recognition \cite{2018Robust}. Adversarial examples could also fool voice-controlled interfaces \cite{2019Imperceptible}, mislead NLP tasks \citep{DBLP:journals/corr/JiaL17}. However, the semantics of binaries limit the applicability of the existing adversarial methods designed against image, voice, or NLP classifiers transplanted to the cybersecurity realm, because there is a structural interdependence between adjacent bytes. \citet{DBLP:journals/corr/abs-1801-08917} introduce one way to bypass machine-learning-based detection by manipulating the PE file format. They find several structures in windows PE files that could be changed without affecting their functionality. \citet{DBLP:journals/corr/abs-1802-04528} craft bytes adversarially in regions that do not affect execution. Specifically, they append adversarial bits at the end of files. \citet{2019Exploring} extend this idea by finding more places to append in PE files, such as in the middle of two sections of PE files. Different padding strategies are also evaluated, including random appending, FGM appending, and benign appending. \citet{2019Optimization} manipulate instructions to produce adversarial examples. Instructions are a functional part of binary files. They introduce two families of transformation. The first one named in-Place randomization (IPR) is quoted from \citep{6234439}. The second one named code displacement (Disp) is also adopted in our article as the baseline.
Disp relocates sequences of instructions that contain gadgets from their original locations to newly allocated code segments with a $jmp$ instruction. \citet{2019Optimization} extend the Disp algorithm. They make it possible to displace any length of consecutive instructions, not only the ones that belong to gadgets.
In terms of the variable space, they focus on the structure characteristic or the code characteristic. None of the above-mentioned articles discussed the data segment, while it plays an important role in malware classification.  \\
\subsection{Explanation of Adversarial Machine Learning in Malware}
In the field of malware, the traditional routine of producing adversarial examples in black-box settings involves proposing function-preserving actions and using a complex or random method \cite{2019Optimization} to bypass DNNs. Many articles introduce different ways to transform malware without changing its functionality, but only a few articles have explained why their approaches work. Due to the nonlinearity of DNNs detector, they rely on unexplainable methods or random ways to optimize their transformation. Unexplainable methods help us little in the design of the malware detector. \citet{2020Fuctionality} use the genetic algorithm to generate the adversarial examples. \citet{2019Optimization} use the transformation randomly. \citet{DBLP:journals/corr/abs-1801-08917} use DNNs based reinforcement learning to evade the detector which is also unexplainable. While DNNs have shown great potential in various domains, the lack of transparency limits their application in security or safety-critical areas. \citet{DBLP:journals/corr/abs-2010-09470} claim that artifacts unrelated to the classified target may create shortcut patterns for separating different classes. Consequently, the DNNs may adapt to the artifacts instead of the original problems. It is important to explore what these models have learned from malware. An explainable technique is needed to tell us the most influential features. Most of the existing researches on the interpretability of DNNs focus on image classification and NLP processing \citep{2016lime}\citep{unknownExplaining}\citep{2017A}. To improve the transparency of malware classification, researchers have started to work on the explanation issue of malware classification. To the best of our knowledge, \citet{coull2019activation} are the first to explore this topic. They use different methods such as hdbscan, shaply value and byte embeddings to analyze the model. They examine the learned features at multiple levels of resolution, from individual byte embeddings to end-to-end analysis of the model. Jeffrey Johns et al. also study what DNNs learned in malware classification by analyzing the activation of the CNN filter. They point out that a CNN-based malware detector could find meaningful features of malware \citep{articletRepresentation}.
Luca Demetrio's work is the closest one to our research. They use Integrated Gradients to find the most important input and point out that the MalConv model does not learn the key information in the PE file's header according to the interpretability analysis \citep{DBLP:journals/corr/abs-1901-03583}. They devise an effective adversarial scheme based on the explanation. \citet{DBLP:journals/corr/abs-2009-13243} make use of the explainable techniques to train a substitute neural network to represent the attacked malware classifier. They attack the substitute model instead. Different from these articles, our function-preserving actions are capable of processing more segments(data). We also do not need a large number of examples to train the surrogate model.
\section{TECHNICAL APPROACH}
In this section, we discuss the technical approach behind our framework. We start by describing the general algorithm. Then, we introduce the approximating boundary model to fit the DNNs. Thereafter, we show the data augmentation module. At last, we explain in detail how we craft adversarial examples. Throughout the paper, we use the following notations. $m$ refers to the original malware, $f(m)$ refers to the output of the DNNs detector (e.g. class probabilities or logits). We use $g(m)=\vec{m}\circ \vec{w}$ to approximate $f(m)$. $\vec{m}$ is the interpretable data representation vector. $\vec{w}$ is the weight of the linear equation. $\tilde{m}^{i}$ refers to the perturbed malware at the $i$-$th$ round of the algorithm and $\tilde{m}^i_{j}$ is the $j$-$th$ perturbed example in each round. $r_{j}$ is the perturbation that we make. 
\subsection{Instance Attack}
In the field of adversarial example of binary code in black-box setting, because of the constraints of semantic inconvenience. The transformation can only be limited to a few discrete actions as mentioned in Section \ref{section:Function Invariant Transformation}. The transformation is discrete and difficult to optimize with gradient-descent alike methods. We exploit the interpretable models to make adversarial attacks. Adversarial attacks can roughly be divided into three categories: gradient-based, transfer-based, and score-based (or decision-based) attacks \cite{brendel2018decisionbased}. Our method is somewhat similar to the transfer-based method. A traditional transfer-based method train a substitute neural network model on a training set that believed to accurately represent the attacked malware classifier such as \cite{DBLP:journals/corr/abs-2009-13243}. To train a surrogate model, traditional methods require a large number of examples. This is often impossible in practice. We need one example. Our framework is instance-based and fits a specific example, not the detection model. That is we use a local Interpretable algorithm to train a substitute model that believed to accurately represent the specific instance. For example, a DNN detector $f$ and a malware $m_{0}$, we can train a substitute model $g(m_{0}^1)=f(m_{0}^1)$,$g(m_{0}^2)=f(m_{0}^2),\cdots $. But for another malware $n$, $g(n_{0})\ne f(n_{0})$. \\
We use a locally generated linear function $g$ to find the importance features of binary, and then we use the function invariant transformation mentioned in \ref{section:Function Invariant Transformation} to remove that features of malware. Because only one binary is required, we name it Instance Attack. 
\subsection{General Framework}
We work in a black-box setting. We assumed that we have no access to classify model and data set. We don't have any idea about the structure of the classify model and the data distribution. We are given only one binary instance. There is nolimitation for the number of queries. We can arbitrarily transform the instance. Our target is to generate a binary to mislead the classifier. After the transformation, we must guarantee the program functionality.\\
The basic intuition of our framework is to approximate the result of the specific example with a linear function and make the perturbation towards the approximate decision boundary iteratively.
The whole procedure works in rounds, where each round consists of three steps. In the first step, we use data augmentation to generate a large number of new samples from the original binary. Then we could use a linear model to fit these samples. 
Here, we use the FastLSM algorithm (described in detail in Section 3.4) to fit the malware detector $f$ with a linear function $g$. The second step is to approximate the decision boundary by solving the linear equations $g$, and transforming the most important part of the malware accordingly to make $g(\tilde{m}^{i})=benign$. For the function invariant restriction of the binary file, we deform $\tilde{m}^{i}$ to $\tilde{m}^{i+1}=\tilde{m}^{i}+r^{i}$ with the function invariant transformation we proposed in Section 3.5. We finally query the black-box detector to get $f\left(\tilde{m}^{i+1}\right)$. If the result of $f(\tilde{m}^{i+1})$ is still malware, let's move on to the next iteration or we will end the algorithm if the maximum number of iterations is reached. Algorithm \ref{algorithm:General algorithm.} presents the pseudocode of the overall procedure. 
\begin{algorithm}[!h]
	\caption{General algorithm}
	\label{algorithm:General algorithm.}
	\hspace*{0.02in}{\bf INPUT:}
	a malware $m$, a classifier $f$, a linear equation $g$, the approximating algorithm FastLSM$\left(\right)$, a functional invariant transforming function Tran$\left(\right)$, address of most weighted data $[start,end]$\\
	\hspace*{0.02in}{\bf OUTPUT:}
	a new malware $\tilde{m}^{i}$
	\begin{algorithmic}[1]
		\While{$i<max iteration$}
		\State $g(\tilde{m}^i) \leftarrow FastLSM(\tilde{m}^i,f())$
		\State $start,end \leftarrow solving \quad g(\tilde{m}^i)=benign \quad $
		\State $\tilde{m}^{i+1} \leftarrow Tran\left(start,end,\tilde{m}^i\right)$
		\If {$f(\tilde{m}^{i+1} ) == malware$} 
		\State $i\leftarrow i+1 $
		\Else
		\State\Return \quad $success$
		
		\EndIf
		
		\EndWhile\\
		\Return \quad $ false$
	\end{algorithmic}
	
\end{algorithm}
\subsection{Formalizing The Model}
Adversarial examples are variants of normal examples by adding some imperceptible perturbations. The adversarial examples make the detection model misclassify examples with high confidence. \citet{2016Towards} model the adversarial examples as a constrained minimization problem:
\begin{eqnarray}
min \quad D\left(\tilde{m}^{i},\tilde{m}^{i+1}\right) \\
s.t \quad f\left(\tilde{m}^{i+1} \right)=benign
\end{eqnarray}
$\tilde{m}^{i}$ is fixed and the object is to find the perturbation $\tilde{m}^{i+1}$ that can minimize $D\left(\tilde{m}^{i},\tilde{m}^{i+1}\right)$ and further lead to the evasion. $D\left(\tilde{m}^{i},\tilde{m}^{i+1}\right)$ is a distance function and the perturbation is subject to $f\left(\tilde{m}^{i+1} \right)=benign$. Since the identifier is a black-box model, it is difficult to find a solution for the original function $f$. \citet{2016Towards} proposed to solve a simple objective functions $G$ instead, $ G\left(\tilde{m}^{i+1} \right)=benign$ if and only if $f(\tilde{m}^{i+1})=benign$.
\begin{eqnarray}
min \quad D\left(\tilde{m}^{i},\tilde{m}^{i+1}\right) \\
s.t \quad G\left(\tilde{m}^{i+1} \right)=benign
\end{eqnarray}
In malware detection, we also use these formulas to find adversarial perturbations $\tilde{m}^{i+1}$ for the original binary $\tilde{m}^{i}$ that target a class $f_{benign}$. In a black-box environment, finding an object function $G$ such that $ G\left(\tilde{m}^{i+1} \right)=benign$ if and only if $f(\tilde{m}^{i+1})=benign$ is an overly strong requirement. Here we use the local explanation $g$ instead. When $f(\tilde{m}^{i+1})=benign$ is true, $g(\tilde{m}^{i+1})=benign$ is established, but the opposite is not necessarily true. When $g(\tilde{m}^{i+1})=benign$ is true, $f(\tilde{m}^{i+1})$ may not be $benign$. The optimization could be converted to the following problem:
\begin{eqnarray}
min \quad D\left(\tilde{m}^{i},\tilde{m}^{i+1}\right) \\
s.t \quad g\left(\tilde{m}^{i+1} \right)=benign
\end{eqnarray}
We can easily solve $Eq.\left(6\right)$ to get the minimal perturbation $r$ from $\tilde{m}^{i}$ to $\tilde{m}^{i+1}$, since $g$ is a linear equation. Because $g(\tilde{m}^{i}+r)$ is the approximation of $f(\tilde{m}^{i+1})$, we must incorporate the perturbation $\tilde{m}^{i+1}$ 
back into the original classifier $f$ to get the accurate value. If $f\left(\tilde{m}^{i+1} \right)=benign$, then we will finish the algorithm or we should recompute the linear approximation of $f\left(\tilde{m}^{i+1} \right)$. We will continue to repeat this process until the evasion is successful or it will reach the maximum number of iterations.   
\subsection{Local Linear Explanations of Malware Detection}
Here, we explain how we build the linear function $g$. Our method is inspired by Local Interpretable Model-Agnostic Explanations(LIME) \cite{2016lime} and Locally Linear Embedding (LLE) \cite{Roweis00nonlineardimensionality}. Deep learning algorithms provide highly satisfactory results. However, their decision procedures are non-linear and we cannot feature out the important parts of the input data directly. When we are given one instance to bypass the classifier, we try to infer how the detector behaves around a specific instance by querying the detector for the results of different transformed examples. We use the data augmentation method to produce the adversarial perturbations. As claimed in \cite{Roweis00nonlineardimensionality}, we assume that binary files can be represented by points in a high-dimensional vector space. The binary and its transformations lie on or close to locally linear path of the manifold. Because coherent structure between the binary and its variants leads to strong correlations, we can characterize them by linear coefficients. We think that the linear model is simple enough and is interpretable.
\begin{equation}
\xi(m)=\operatorname{argmin}_{g \in G} \mathcal{L}\left(f, g, \Pi_{m}\right)+\Omega(g)
\end{equation}
$Eq.\left(7\right)$ describes how to solve this problem, where $f$ denotes a DNNs detector, and $g$ is an explainable model to approximate $f$ without knowing its parameters. In classification, $f\left(m\right)$ produces the probability (or a malware indicator) that $m$ belongs to a certain category. If $g$ is a potentially interpretable function, and $\Omega\left(g\right)$ measures the complexity of the explanation. $\mathcal{L}\left(f, g, \Pi_{m}\right)$ measures how unfaithful $g$ is in approximating $f$ in the locality defined by $\Pi_{m}$. To ensure the interpretability and local closeness, $\mathcal{L}\left(f, g, \Pi_{m}\right)$ should be minimized and $\Omega\left(g\right)$ should be low enough. The lower the $\Omega\left(g\right)$, the easier it is for humans to understand the model.
In this paper, $G$ could be the class of linear models, such that $g\left(m\right) =\vec{w_{g}} \circ  \vec{m}$. We can get the interpretable representation $\vec{m}$ of $m$ directly, the specific rule will be described in the following section. We define $\Pi_{m}\left(\tilde{m}^i_{i} \right)=D\left(\tilde{m}^i_{i},\tilde{m}^i_{0}\right)$, where $D$ is some kind of distance function, for example, the $L2$-norm distance. $\tilde{m}^i_{0}$ is the specific example and $\tilde{m}^i_{j}$ is the perturbed example in each round.  We carefully choose some perturbed examples $\tilde{m}^i_{1},\tilde{m}^i_{2},\tilde{m}^i_{3} \cdots ,\tilde{m}^i_{j} $ within each round, the method of perturbed, see Section \ref{section:Data Augmentation Module}. $\mathcal{L}$ could be a locally weighted square loss as defined in $Eq.\left(9\right)$. In this way, the function has been converted into a linear function fitting problem.  
\begin{eqnarray}
g\left(\tilde{m}^i_{j}\right)=\vec{w_{i}} \circ \vec{m}^i_{j}\;\\
\mathcal{L}\left(f, g, \Pi_{m}\right)=\sum_{j}\; \Pi_{m}(\tilde{m}^i_{i})\left(f(\tilde{m}^i_{j})-g\left(\tilde{m}^i_{j}\right)\right)^{2}
\end{eqnarray}
Given a malware instance $\tilde{m}^i$ and $f$, this problem is a typical ordinary least square problem. We sample examples around $\tilde{m}^i$ by drawing non-zero elements uniformly at random to get the value of $\tilde{m}^i_{j}$. $f(\tilde{m}^i_{j})$ could be obtained by querying the classifier. $\vec{m}^i_{j}$ is the interpretable data representation that can be easily calculated. By solving these functions, we can get the weight $\vec{w_{i}}$. We call $g$ the local explanation model of $f$. We could also use $g$ as the approximate function of $f$.
\subsection{Data Augmentation Module And Optimization Algorithm}
\subsubsection{Data Augmentation Module}
\label{section:Data Augmentation Module}
The ablative analysis is often used in evaluating the DNNs model. It is a technique for evaluating machine learning models by discarding certain features\cite{Ablation2}. We adopt similar ideas, but our model is instance-based. We are not to quantify the DNNs detector, but the important portion of a specific instance.
We make augmentated examples from a specific instance by discarding (masking) certain features. Then we use a linear function $g$ to fit the DNNs around this examples, and we calculate the weights for each feature. 
Computing the weight of all bits is time consuming. Given a file of length $LM$, it would take $o(LM^3)$ time to find the most important bit using the Least Square Method. To improve efficiency, we propose two optimization mechanisms. 
First, we use the notion of the superpixel as the basic unit of interpretable representation. Then, we introduce the FastLSM algorithm to reduce computational complexity.
\subsubsection{Interpretable Data Representations and Segmentation Algorithm}
\begin{figure}[H]
	\centering
	\includegraphics[width=.79\columnwidth]{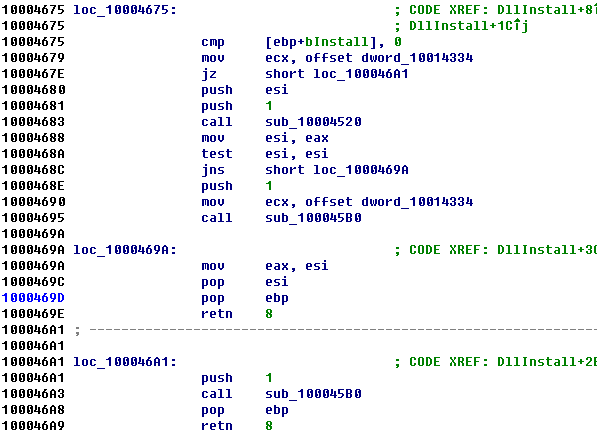}
	\caption{An Example of Superpixels.}
	\label{fig:Examples of Super pixels.}
\end{figure}

Superpixels were used in image segmentation originally \cite{DBLP:journals/corr/abs-1907-06119}. Common image segmentation algorithms include quick shift, felzenszwalb, slic. In this paper, superpixels are the results of a binary file over segmentation. We could use the tools to disassemble the binary file, such as IDA, Binary ninja. Superpixels are the basic function blocks returned by disassembly tools. We used the Capstone disassembler framework in our experiment \cite{cap}. A basic block is a straight-line sequence of codes with only one entry point and only one exit. Or we could just segment the examples by their offset. For example, we could divide a 200KB file into ten parts where the size of each part is 20KB. As shown in Figure \ref{fig:Examples of Super pixels.}, there are 3 superpixels returned by the disassemble tool. The start offsets of these superpixels are $0x10004675h$, $0x1000469Ah$, and $0x100046A1h$. The lengths of the three superpixels are $0x25h$, $0x7h$, and $0x8h$ respectively. 
Interpretable data representations $ \vec{m}$ and data features $m$ are different. Features $m$ in range $m\epsilon R^{^{L}}$ are the ground truths. An interpretable representation $\vec{m}$ is a binary vector indicating the ``presence" (denoted by \textbf{1}) or ``absence" (denoted by \textbf{0}) of a patch of codes and its range is $\vec{m}  \epsilon \left\{0,1\right\}^{l}$. $L$ is the length of the features and $l$ is the length of the interpretable representations. Given a file whose content is ``0x1122", ``0x11" ``0x22" are the two super pixels and ``11" is its interpretable representation.  $f(``0x1122")$ equals to $g(``0x1122") =w_{1}*1 +w_{2}*1$. When we sample around ``0x1122", we can get the corresponding interpretable representation. For example, if we transform the ``0x1122" to ``0x0022", its interpretable representation will be transformed to ``01". Table 1 shows the examples in detail.
\begin{table*}[h]
	\begin{center}
		\begin{tabular}{ccccc}
			\toprule
			Interpretable Representation&00&01&10&11\\
			\midrule
			Original Feature&0x0000&0x0022&0x1100&0x1122\\
			\bottomrule
		\end{tabular}
	\end{center}
	\caption{An example of interpretable data representation.}
	\label{tab:An example of interpretable data representations}
	
\end{table*}
\subsubsection{Fast Least Square Method}
\begin{figure}[H]
	\centering
	\includegraphics[width=.79\columnwidth]{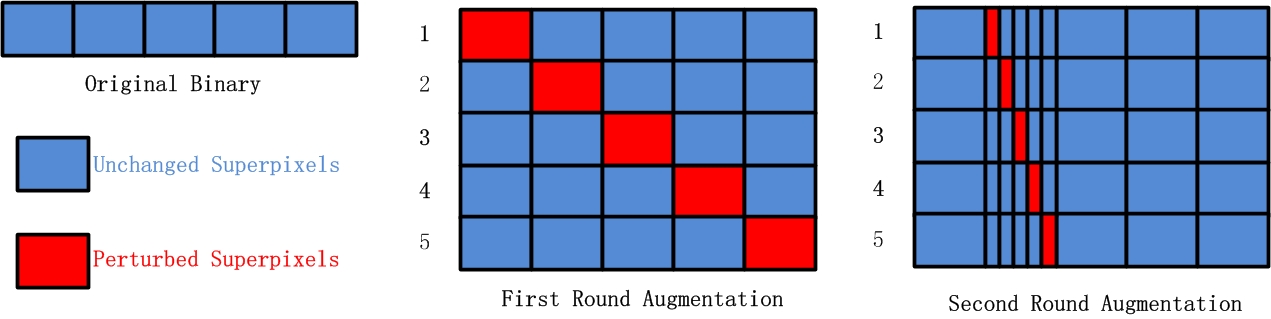}
	\caption{The data augmentation algorithm used in our experiments.}
	\label{fig:The data augmentation algorithm in our experiments.}
\end{figure}
In each round of Algorithm 1, we need to find the minimal perturbation. The weight of the top-$k$ important superpixels rather than each superpixel is needed sometimes. \citet{DBLP:journals/corr/abs-1806-04773} introduce a segmentation algorithm to find the most important superpixel. We extend it and combine it with our local explanation algorithm. We name it the Fast Least Square Method (FastLSM), because it can reduce the computational complexity to $o(log(LM))$ to calculate the weight of the most important superpixel.\\
The steps of FastLSM are as follows. We select the whole binary as the basic superpixel. Firstly, we divide the basic superpixel into $n$ different superpixels. We occlude each super pixel respectively with zero to generate variants and each variant is analyzed by the DNNs $f$. Secondly, we choose the variant that results in a larger drop in classification confidence. Thirdly, if the length of the superpixel with a larger drop is smaller than a specific value $\beta$, the algorithm ends otherwise we will set it as the basic superpixel and go to the first step. We can also occlude the file with random value, null value, or adversarial value \citep{DBLP:journals/corr/abs-1806-04773}. We will discuss how we determine the hyperparameter $\beta$ in Chapter 4.6. Algorithm \ref{algorithm:FastLSM} outlines the basic steps of FastLSM.
\begin{algorithm}[H]
	\caption{FastLSM algorithm}
	\label{algorithm:FastLSM}
	\hspace*{0.02in}{\bf INPUT:}
	a malware $m$, 
	a classifier $f$, a target occlusion size $\beta$, $n$ is the number of segments within each iteration\\
	\hspace*{0.02in}{\bf OUTPUT:}
	a new malware $\tilde{m}$
	\begin{algorithmic}[1]
		
		\State Split file $m$ into $n$ sections, $splitsize\leftarrow \left| m{\div}n\right|, $size of $ith$ section is$\left| m(i)=splitsize\right| $ 
		\State Use LSM to get the weight for each section
		
		\State Find the $jth$ section with the maximum weight
		\State $start \leftarrow$ start address of $jth$ section
		\State $end \leftarrow $ end address of $jth$ section
		
		\While{$ splitsize>\beta$}
		\State $max\leftarrow 1$,$i \leftarrow 1 $

		\While{$i<n$}
		\State $starttemp \leftarrow (start+splitsize*(i-1))$
		\State $endtemp \leftarrow (start+splitsize*i)$
		\State $m_{i}$[starttemp:endtemp]$\leftarrow$ 0x00, occlusion the ith segment with $0x00 $
		\State $starttemp \leftarrow (start+splitsize*(max-1))$
		\State $endtemp \leftarrow (start+splitsize*max)$
		\State $m_{max}$[starttemp:endtemp]$\leftarrow$ 0x00, occlusion the most weight segment with $0x00 $
		\If {$f(m_{i})<f(m_{max})$} 
		
		\State $max \leftarrow i$
		\State$ i\leftarrow i+1$
		
		\Else
		\State$ i\leftarrow i+1$
		
		\EndIf
		\EndWhile 
		\State $start \leftarrow start+splitsize*(max-1)$.
		\State $end \leftarrow start+splitsize*(max)$.
		
		\State $splitsize \leftarrow splitsize {\div}n$
		\EndWhile \\
		\Return $\tilde{m}[start:end]$
	\end{algorithmic}
\end{algorithm}
\subsection{Function Invariant Transformation}
\label{section:Function Invariant Transformation}
The semantics of binaries hinder the direct transplantation of existing traditional adversarial learning methods. Even changes as small as one bit can disturb the original syntax of the binary and may lead to malfunction or even cause the file to crash. For example, if we change 53 to 52 in the binary file, at the assembly level it means that \textbf{push ebx} is changed to \textbf{push edx} and the function of the generated adversarial example would be invalid. Due to the function preserving constraint, the transformations that we use are limited to ones that keep the binary's function intact. In this paper, three families of transformations will be used. The first transformation is an appending algorithm that applies the evasion by appending adversarial bits at the end of the original files \cite{2019Exploring}. These bytes cannot influence the semantic integrity. Appending bytes to unreachable regions of the binary may be easy to detect and could be sanitized statically \cite{2019Optimization}. 
The second transformation that we will use is named code displacement (Disp), which is proposed by \citep{2016Juggling} to break the gadgets originally. \citet{2019Optimization} adopt it to mislead DNNs-based classifiers. The general idea of Disp is to move some codes to the end of the binary. The original codes are replaced with a $jmp$ instruction at the beginning of the code segment. After filling the newly allocated code segment with the adversarial codes, another $jmp$ instruction is appended immediately. 
The third transformation is an original one named data displacement (DataDisp). Disp can only transform code segments rather than data segments. However, we find that sometimes the DNNs attach importance to data segments as shown in the experiment (Section 4.2). So we propose an algorithm that could be used to transform data segments (this transformation can also be applied to code segments with some modifications). DataDisp is a practical code diversification technique for stripped binary executables. DataDisp transfers data to the end of the file and uses the $mov$ instruction to move the data back before the binary is executed. It starts by adding a new section at the end of the PE file. The original codes to be displaced are replaced with adversarial data \citep{DBLP:journals/corr/abs-1806-04773}, random data, or null data.
After filling the newly allocated segment with the $mov$ code, a $jmp$ instruction is appended immediately to deliver control back to the original binary. At last, we will change the OEP to the beginning of the newly allocated section. There are some tips to be noted. If the displaced codes contain important structural information (e.g., \textbf{edata} section) we leave them alone.
\begin{figure}[H]
	\centering
	
	\includegraphics[width=.7\columnwidth]{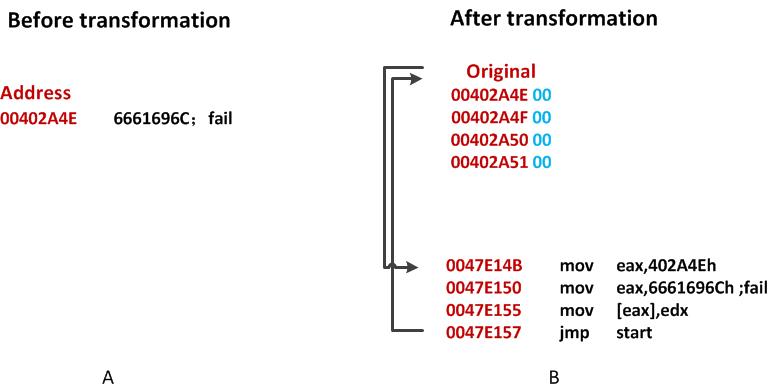}

	\caption{An illustration of DataDisp transformation.	}
	\label{fig:An example of datadisp transformation.}
\end{figure}

As shown in Figure \ref{fig:An example of datadisp transformation.}a, the original data at $0x402A4E$ is string $fail$, and will be replaced by $0x00$. The Original Entry Point (OEP) address is displaced to $0x47E14B$ in Figure \ref{fig:An example of datadisp transformation.}b. Then the address $0x402A4E$ is saved to $EAX$. The new code will reconstruct the original file by $mov$. After the reconstruction, it will $jmp$ back to the original OEP.
\section{EVALUATION}
In this section, we show comprehensive empirical evidence of our adversarial theme. First we will describe the details of the datasets and the DNNs detectors that will be used in our experiments. We also discuss the results of the interpretable analysis on the experimental data. Then we exhibit different transformations. Finally, we compare our model with three different methods. The results illustrate that our adversarial examples generation model is trustworthy. 
\subsection{Datasets and Malware Detector}
\begin{table}[h]

	\begin{center}
		\begin{tabular}{ccccc}
			
			\toprule
			
			Class&All&Train&Validation&Test\\
			\midrule
			Malware&10868&8708&1080&1080\\
			Benign&9814&7814&1000&1000\\
			\bottomrule
		\end{tabular}
	\end{center}
	\caption{The number of binaries for training, validating and testing the DNNs.}	
	\label{tab:the number of benign and malicious binaries for training}
\end{table} 
All experiments were conducted on an Ubuntu 18.04 server with an Intel Xeon CPU and 64GB of RAM. The computer was equipped with Python 3.7, PyTorch, and an NVIDIA GTX3090 Graphics processing unit. We used a mixed data set. We resorted to a publicly available dataset to collect malware. The malware binaries were adopted from the Kaggle Microsoft Malware Classification Challenge \citep{ronen2018microsoft}. 
\begin{table}%
	
	\begin{center}
		\begin{tabular}{cccc}
			\toprule
			\multirow{2}{*}{Classifier}	&\multicolumn{3}{c}{Accuracy}\\
			&Train&Validation&Test\\
			\midrule
			MalConv&98.8\%&97.8\%&96.1\%\\
			AvastNet&99.8\%&97.2\%&97.5\%\\
			
			\bottomrule
		\end{tabular}
	\end{center}
	
	\caption{The DNN's performance}
	\label{tab:The DNN's performance}
\end{table}
This dataset contains nine different malware families. Even though there are no benign ones, we still used the dataset to train our model based on the following consideration. First, most prior works \cite{2018arXiv180404637A}\cite{krl2018deep}\cite{article} used proprietary datasets and some other public datasets present only packed data \cite{208119}. \citet{noever2021virusmnist} contain datasets that contain only processed data. Although \citet{9474321} also offer sufficient raw files, they do not provide benign files either, for copyright reasons. Second, over fifty articles had cited this dataset \cite{ronen2018microsoft}, which was the de facto standard in malware classification. 
We collected benign binaries from a newly created Windows 7 machine. Specifically, we used the \textbf{Portable apps} and \textbf{360 package manager} to install 180 different packages. To make our dataset more representative, we collected popular files (such as Chrome, Firefox, notepad++). We also collected packages that were likely to be used only by academics ( e.g.,MiKTeX, Matlab), developers(e.g.,VsCode, PyCharm), document workers(e.g., WPS, AdobeReader). We selected 9,814 benign binaries, most of them were smaller than 2MB and we divided them into train, validation, and test sets as shown in Table \ref{tab:the number of benign and malicious binaries for training}.\\%
Among all the DNNs classifiers mentioned in Section 2, we selected two classifiers. We trained two different DNNs-based malware detectors with our datasets, based on the representativeness and code availability considerations. Both of the detectors receive raw byte binaries. The first detector is given by \cite{krl2018deep} with four convolution layers and four fully connected layers. It receives inputs of up to 512KB. We would name it AvastNet. The second DNN model is proposed by \cite{article}, the network structure they adopted consists of two 1-D gated convolutional windows with 500 strides. It receives inputs of up to 2MB. We would name it MalConv in the following. We split all files into three sets, train, test and validation. Both of these DNNs achieved accuracy above 95\% on our datasets. The classification results of these two DNNs are reported in Table \ref{tab:The DNN's performance}. \\
As there are no PE headers in the Microsoft dataset, we could not disassemble these binaries to validate our algorithm. We resorted to \textbf{VirusShare} to download malware belonging to the nine malware families that had not appeared in the Microsoft dataset. VirusShare is an open repository of malware examples that contains labels. We downloaded 88 binaries that did not appear in the training set, but belonged to the same nine families. We validated them with MalConv and AvastNet separately. 68 of them were marked as malware in MalConv as shown in Table \ref{tab:The number of benign and malicious binaries for test2}. We also sampled 88 benign binaries for the test, 10 of them were misclassified in MalConv. 
72 of them were marked as malware in AvastNet as shown in Table \ref{tab:The number of benign and malicious binaries for test2}. We also sampled 88 benign binaries for testing, 8 of them were misclassified in AvastNet. 

\begin{table}[h]
	\begin{center}
		\begin{tabular}{ccccc}
			\toprule
			Class&Model&all&as malicious&as benign\\
			\midrule
			Malware&MalConv&88&77.3\%&22.7\%\\
			Benign& MalConv&88&11.4\%&88.6\%\\
			Malware&AvastNet&88&81.2\%&18.8\%\\
			Benign&AvastNet&88&9.1\%&90.9\%\\
			\bottomrule
		\end{tabular}
	\end{center}
	\caption{Binaries used to test our model}
	\label{tab:The number of benign and malicious binaries for test2}
\end{table}

\subsection{Weight Analysis On Superpixels}

\subsubsection{Distribution }

\begin{figure}[h]
	\centering
	\includegraphics[width=.6\columnwidth]{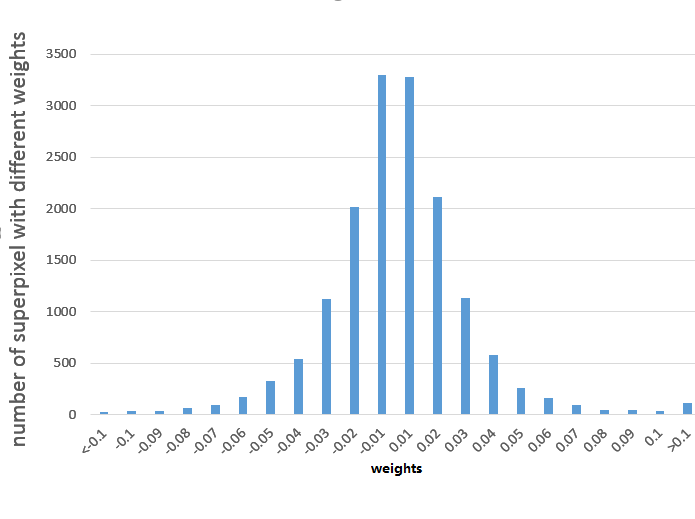}
	\caption{Weight distribution diagram under the MalConv classifier. It shows that the curve of the weight distribution of malware follows the normal distribution.
	}
	\label{fig:The distribution diagram of weight.}
\end{figure}

\begin{figure*}
	\centering
	\subfloat[Positive Data]{
		\begin{minipage}{7cm}
			\centering
			\includegraphics[scale=0.45]{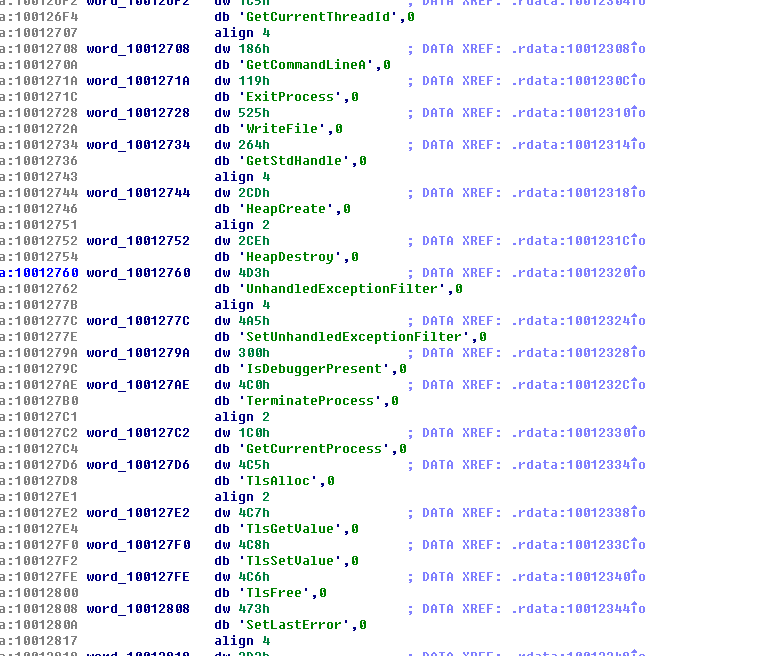}
		\end{minipage}
	}
	\subfloat[Negative Data]{
		\begin{minipage}{7cm}
			\centering
			\includegraphics[scale=0.7]{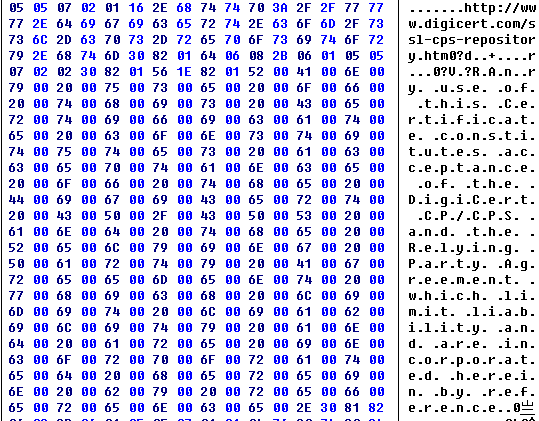}
		\end{minipage}
	}
	\caption{Data with different weight. The data on the right (b) have a negative impact on the results. The data on the left(a) have a positive impact. 
	}
	\label{fig:Data with different weight.}
	
\end{figure*}

\begin{table}[h]

	\begin{center}
		\begin{tabular}{ccccc}
			\toprule
			Weight&$\left[ 0,0.01 \right]$ & $\left[ 0.01,0.1 \right]$ & $\left[ 0.1,0.2 \right]$ &$\left[ 0.2,1 \right]$\\
			\midrule
			MalConv&43.2\%&50.8\%&5.5\%&0.5\%\\
			AvastNet&44.2\%&50.2\%&5.2\%&0.4\%\\
			\bottomrule
		\end{tabular}
	\end{center}
	\caption{Proportion of superpixels with different weight}
	\label{tab:Proportion of blocks with different weights}	
\end{table}
Using the least square algorithm, we could get the weight of different parts of the binaries. Through the statistical analysis, we gained some understanding of the black-box classifier. We divided the sampled files into superpixels that are smaller than 8KB and larger than 4KB. We summarized the corresponding weight of these superpixels on the MalConv\&AvastNet classifiers. As shown in Table \ref{tab:Proportion of blocks with different weights}, the weight of 90\% superpixels was less than 0.1, only about 5\% of the superpixels were between 0.1 and 0.2. 0.5\% of the superpixels had a weight greater than 0.2. We can even estimate their distribution. We just plot the weight of different superpixels of the malware under MalConv, as shown in Figure \ref{fig:The distribution diagram of weight.}. In general, although the weight sums of all superpixels must be greater than 0, the weight distribution conforms to the normal distribution, and the mean is about zero. Figure \ref{fig:The distribution diagram of weight.} shows the result of weight analysis on MalConv, the analysis on AvastNet looks alike.\\
According to the weight distribution diagram, we conclude that most of the contents do not have much impact on the result. There are only less than one percent of the superpixels that have a significant impact on the classifier,whose weight is greater than $0.2$. We could use these data as adversarial data in the following experiments. Some superpixels with higher absolute weight are listed below. The data that is shown in Figure \ref{fig:Data with different weight.}b has a negative impact on the MalConv classifier. It's a URL for \textbf{digicert.com}. Obviously, this website is not malicious. The existence of these codes in binaries can increase the probability of being judged as benign. The data in Figure \ref{fig:Data with different weight.}a plays positively to AvastNet, this figure is the disassembly result of IDA Pro. We can see that the code is the \textbf{import table} of a PE file. Many of the functions in the table are related to malicious behavior with high probabilities, such as the \textbf{isdebuggerpresent} which is the API that is often used by malware to resist reverse analysis.\\

We also analyzed the malware that contains malicious APIs with the lightGBM model which was introduced by \citep{2018arXiv180404637A}. The lightGBM model was not trained on the same dataset as our model. Although the output of lightGBM was 0.8388964 which implied that there was a great chance that the file was malware. But by analyzing the file with our interpretable model, we can see that the model gave too much weight on the file's PE header which was not in accordance with the experience \cite{8553214} and it could be evaded easily \citep{DBLP:journals/corr/abs-1901-03583}. We show the weight and offset of the three most weighted superpixels in Table \ref{tab:The Top Five segments of lightGBM model on the example}. We could conclude that the lightGBM model makes decisions according to false causalities. \\
\begin{table}[h]
	
	
	\begin{center}
		\begin{tabular}{cccc}
			\toprule
			Offset&0x0000-0x1000&0xe000-0xf000&0xc000-0xd000\\
			\midrule
			Weight&0.93&-0.0277&-0.0272\\
			\bottomrule
		\end{tabular}
	\end{center}
	\caption{The weight and offset of the three most weighted superpixels of the malware under the lightGBM. We concluded that the lightGBM model make decision according to false causalities.}
	\label{tab:The Top Five segments of lightGBM model on the example}
\end{table}
\subsubsection{Proportion of Code Segment Weight}

\begin{figure}[H]
	\centering
	\includegraphics[width=.6\columnwidth]{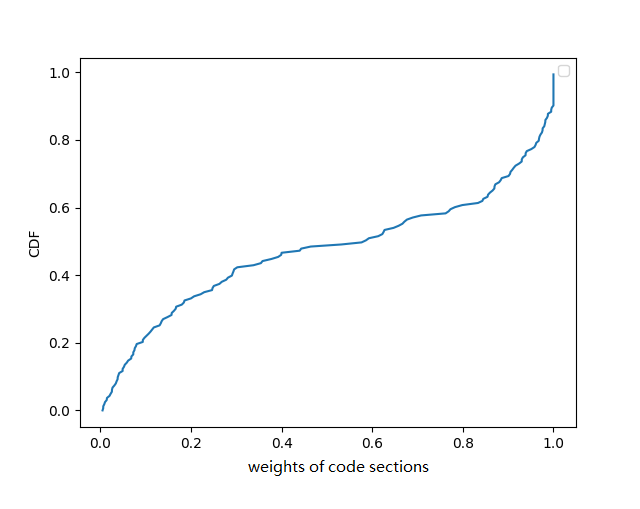}
	\caption{A CDF graph of code sections' weight.
		\label{fig:An CDF graph of weights of code sections.}
	}
	
\end{figure}

We also studied the proportion of code segments in the overall score returned by the classifiers. We used the explanation-based model to analyze the result. We calculated the weight of the code segments by adding the weight of all the superpixels belonging to the code segments. Although it was not strictly defined, it corresponded to the code/text section of the binaries \cite{c22}. But the writer of the malware could change the name of the code segment. By using our explanation model, we could get the weight of all the sections (bss, edata, idata, idlsym, pdata, rdata, reloc, rsrc, sbss, sdata, srdata, code/text). We calculated the weight of code/text for all binaries. The CDF result was shown in Figure \ref{fig:An CDF graph of weights of code sections.}. We could see that the weight of the code sections is about 50\% in half of all the binaries. Although this was only an estimation, the weight of the code segments must be limited. We concluded that code sections only account for part of the weight. Because all the data segments could not be transformed by Disp, there is a high probability that the success rate will be limited if we only use the Disp algorithm. Because the DataDisp algorithm targets data variable spaces, it could achieve a better result than Disp.

\subsection{Randomly Applied Transformations}

In order to study the influence of the location of the transformed content and the type of the transformed content on the success rate of adversarial examples, we evaluated whether the randomly applied transformation would lead to evading the DNNs. For each binary that we used to evaluate the transformations, we created up to 200 variants. If the detection results of more than one variants changed, we would consider the transformation successful.  We divided the binary into superpixels. For code sections, superpixel was the basic function block returned by disassemble tools. For data sections, we divided them into 1KB length superpixels by offset. The so-called random applied transformations had two meanings. First, for a certain superpixel within a binary, whether to transform it was random. Second, the gap space after the transformation was filled with adversarial data or random data. Specifically, we designed two experiments. In the first experiment, we used random data to fill in the blanks and in the second experiment, we used adversarial data to fill in the blanks. The adversarial data was the data we found in the previous section with a high absolute weight. \\
We ran the two experiments with the setting that increased binaries' size under 5\% and within 200 iterations for Disp and DataDisp. When we used Disp \&
DataDisp randomly with random data, the result was that 3 malware were misclassified and 5 benign binaries were misclassified as malware for MalConv. 4 malware and 6 benign binaries were misclassified for AvastNet. The results are easily explained under our framework, because the weight is under a normal distribution with a mean value of 0 as shown in Figure \ref{fig:The distribution diagram of weight.}. If we use the Disp\&DataDisp algorithms randomly, the weight of transformed binaries is also under a normal distribution and the sum of the weight has a high probability with a mean value of 0. There is a high probability that the adversarial examples will not evade the detector. So we could conclude that it's not the DNNs are robust against naive Disp transformations as claimed in \citep{2019Optimization} but it's just a matter of probability. When we used adversarial data to fill in the gaps, the results were much better. These adversarial data came from the higher weighted data we found in the previous section. We got the highest success rate of 24\% for MalConv and 39\% for AvastNet.

\subsection{Evaluation On Transformations: Disp and DataDisp}

In this section, we evaluated the Disp and DataDisp transformations respectively. We used the interpretable model to optimize the procedure. For Disp, we set 5\% as the maximum displacement budget and 200 as the maximum number of iterations. We filled the gap left by the transformation with adversarial data. The Disp algorithm could achieve a max success rate of 59\% for MalConv and 45\% for AvastNet. For the DataDisp algorithm, we set 5\% as the maximum displacement budget and 200 as the maximum number of iterations. We filled the gap left by the transformation with adversarial data. The DataDisp algorithm achieved a max success rate of 53\% for MalConv and 35\% for AvastNet.\\
\citet{2019Optimization} also tested Disp with a hill-climbing approach. They only moved subsections that had a positive impact on the results. They got a maximum success rate of 24\% . We thought that it was because Disp could only transform the code section of a binary file.

\subsection{Evaluation On Explanation-Based Adversarial Algorithm}

\begin{figure*}
	\centering
	
	\subfloat[MalConv]{
		\begin{minipage}{7cm}
			\centering
			\includegraphics[scale=0.45]{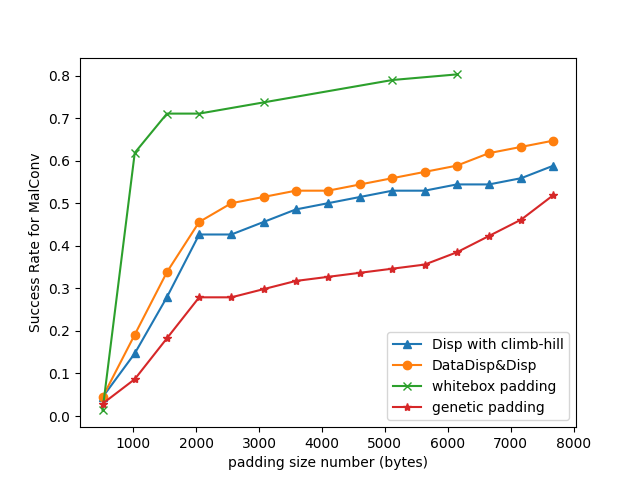}
		\end{minipage}
	}
	\subfloat[AvastNet]{
		\begin{minipage}{7cm}
			\centering
			\includegraphics[scale=0.45]{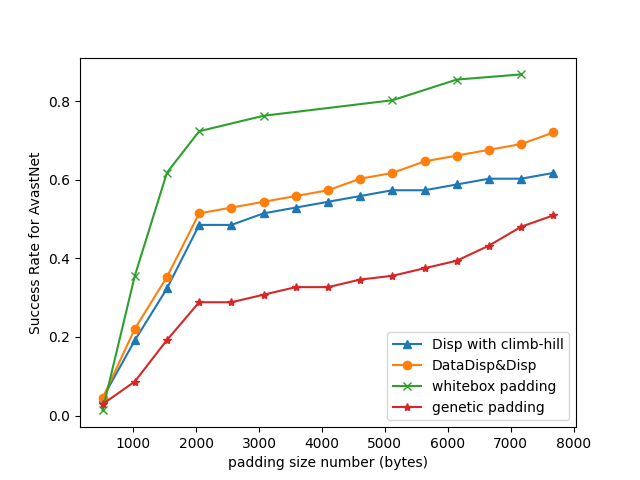}
		\end{minipage}
	}
	
	\caption{A Demonstration of different attacking algorithms. The orange line is the misclassified rate of our black-box algorithm which is not as effective as the gradient-based algorithm and is better than genetic padding and Disp with climb-hill algorithm within a certain range.
	}
	\label{fig:An demonstration of different padding strategies.}
	
\end{figure*}

To assess our explanation-based model, we compared our model with other algorithms in this subsection. We set 5\% as the maximum displacement budget and 200 as the maximum number of rounds and we used a combination of Disp \& DataDisp transformations. We compared our model with three different model, they were Disp with a hill-climbing approach \cite{2019Optimization}, genetic padding \cite{2020Fuctionality} and gradient-based attack \cite{DBLP:journals/corr/abs-1802-04528}. 
All these algorithms would increase the length of the binary, because different content would be padded at the end of the file. The gradient-based algorithm worked in a white-box setting, they used the parameters of DNNs to calculate the gradient \citep{2019Exploring}\cite{DBLP:journals/corr/abs-1802-04528}. The gradient-based padding we used was adapted from \cite{DBLP:journals/corr/abs-1802-04528} with epsilon 0.5 and iteration 2. The genetic padding was a black-box algorithm and was adapted from \citep{2020Fuctionality} with iteration 10 and population 50. The genetic padding needed data randomly sampled from different files. Our algorithm also worked in the black-box setting and we added the binary files with the most weighted data that we found in the previous section. The result is shown in Figure \ref{fig:An demonstration of different padding strategies.}. 

Although our algorithm was not as effective as the gradient-based model, we could see that our method was better than genetic padding and Disp with the hill-climbing approach. The gradient-based attack needs model information that is unavailable in practice. Our attack model had the best performance of all the black-box models. While without budgets restriction, our explanation-based model caused more binaries to be misclassified. 
\subsection{Miscellaneous}

$\textbf{Hyperparameters}$. Throughout this article, we chose 200 as the maximum number of iterations and 5\% as the maximum displacement budget, we also used 1KB as the size of the superpixel of DataDisp in section 4.3. We used these hyperparameters because our method could achieve almost perfect success under this configuration. In \cite{2019Optimization}\cite{10.1145/3433210.3453086}, they tested Disp with a hill-climbing approach with similar hyperparameters.

$\textbf{Integrity of binary}$. To ensure the functionality of binaries was intact after transformation. First, we selected 6 different binaries and manually checked their instructions with OllyDbg \cite{ollydbg}. Second, we chose 10 different benign binaries and manually checked their functionality by running them on Windows. All the file functions well. Third, we also used the Cuckoo Sandbox \cite{cuckoo} to test 10 malware. One of them collapsed after transformation, and the rest function normally. We check the file manually. We figure out that the binary does not strictly follow the PE format specification. The length of data segment shown in fileheader does not match the actual length.

\section{DISCUSSION AND FUTURE WORK}
In this section, we discuss the results of the experiments and the improvement that we plan to investigate in the future. We also give a brief discussion on the limitation and the foundational assumption of our model.

\subsection{Discussion On Experiments}

We carefully designed multiple sets of experiments. First, through the explanation analysis of binary files, we could find out what the black-box detector valued on a specific binary and we could make targeted attacks. Second, we tried the transformations randomly in section 4.3. We found that randomly applied transformations could not achieve high adversarial results. It illustrated that both optimization and adoption of adversarial data were important. Third, we compared our model with others. Our model didn't perform as well as the gradient-based attack model. It was easy to understand, the gradient-based model worked in white-box settings. The gradient-based attack needs model information which is unavailable in practice. Our adversarial model could transform both data and code segments with Disp \& DataDisp transformation methods. It performed the best under black-box conditions.
\subsection{Discussion On The Model}
When attacking black-box systems, adversaries do not know the model's internal information. Adversaries could learn sufficient information to recover the classifier, which is named the adversarial classifier reverse engineering (ACRE)\cite{10.11451081870.1081950}. Or they could train a substitute model using synthetic inputs generated by adversaries\cite{DBLPjournalscorrSzegedyZSBEGF13}. They assume that two models solving the same ML task with comparable performance are likely to have similar decision boundaries\cite{8406613}. We make the similar assumption that two models with comparable performance around one instance are likely to have similar decision boundaries. So we train a substitute model around a specific example. Compared with training a substitute model on the whole set of data, the instance-based approach greatly reduces the computational complexity of training and reduces the amount of data required.\\
In our model, only one training instance is needed instead of many. We often only get a few specific samples in reality, and we need to generate adversarial samples in a targeted manner. All the inputs to train the substitute model are transformed from the instance itself. Because coherent structure between the binary and its variants leads to strong correlations. We believe that the information learned from perturbed instances is specific and targeted. We can train a simpler substitute model and attack this model instead. We name it instance-based attack. In Locally Linear Embedding(LLE) \cite{Roweis00nonlineardimensionality}, each data point is a linear combination of its neighbors. 
As claimed in LLE \cite{Roweis00nonlineardimensionality}, we assume that the binary and its perturbed instances lie on or close to a locally linear path of the manifold. So we can characterize each binary from its neighbors by linear coefficients. This is the key assumption of our article. Although we use an interpretable linear function to fit the classifier in our model, it is not limited to this. Even if the binary and its perturbed instances do not lie on a locally linear path, the structures between the binary and its variants are still highly relevant. In this paper, we use an interpretable model that is simpler enough to be understood by humans. We could also train a complex substitute fitting model but is simpler than the original DNN model, such as a local non-linear model \cite{Guo2018LEMNAED} or a DNN model with less parameters.\\
As we have discussed the mathematical basis of the instance attack model, we figured out that our fitting function is a necessity and not a sufficient one. We resolve to iterate many times. In contrast to NLP and image classification, minimizing the perturbation is not the most important restriction but keeping function intact. There may be no similarity in the characters of two semantic invariance binaries. It is the reason why we could make so many transformations. 
\subsection{Limitations and Future Works}
Limited by our linear fitting model, our interpretable model is not suitable for some structure-based adversarial transformations such as content shifting \cite{10.11453473039}\cite{DBLP:journals/corr/abs-1801-08917}. Our algorithm is instance-based which is to do a fitting for each example, it needs a lot of queries and calculations. But the convergence of the algorithm has not been proved. We use a linear function to fit the classifier. We believe that we could introduce some more complex models such as the local non-linear interpretable model \citep{Guo2018LEMNAED}, and the accuracy can be furtherly improved. The combination of global fitting and local fitting frameworks is also worth studying and the intrinsic dimension of our model could also be discussed \cite{pope2021the}.
\section{CONCLUSIONS}

In this paper, we introduce the notion of the instance-based attack. We analyzed two DNNs based malware classifiers with our interpretable model. We got some characteristics of the two models under black-box conditions. We found that data segments play an important role in the results. None had discussed the importance of data segments in related articles. We introduced a novel method to generate adversarial examples, we name it instance attack. Different from other methods which insert code in invalid places or transform only code segments, our adversarial model could make transformations on both the data and code segments with Disp \& DataDisp. Our model can achieve state-of-the-art results under black-box conditions. Most importantly, instance attack can be verified with domain knowledge. We hope our work will be an inspiration for others.
\section*{ACKNOWLEDGMENTS}
We would like to thank the anonymous reviewers for their valuable comments. This work was support by National key R\&D Plan:2018YFB0805000.




\bibliographystyle{IEEEtranN}
\bibliography{sigproc.bib}

\begin{thebibliography}{51}
\providecommand{\natexlab}[1]{#1}
\providecommand{\url}[1]{#1}
\csname url@samestyle\endcsname
\providecommand{\newblock}{\relax}
\providecommand{\bibinfo}[2]{#2}
\providecommand{\BIBentrySTDinterwordspacing}{\spaceskip=0pt\relax}
\providecommand{\BIBentryALTinterwordstretchfactor}{4}
\providecommand{\BIBentryALTinterwordspacing}{\spaceskip=\fontdimen2\font plus
\BIBentryALTinterwordstretchfactor\fontdimen3\font minus
  \fontdimen4\font\relax}
\providecommand{\BIBforeignlanguage}[2]{{%
\expandafter\ifx\csname l@#1\endcsname\relax
\typeout{** WARNING: IEEEtranN.bst: No hyphenation pattern has been}%
\typeout{** loaded for the language `#1'. Using the pattern for}%
\typeout{** the default language instead.}%
\else
\language=\csname l@#1\endcsname
\fi
#2}}
\providecommand{\BIBdecl}{\relax}
\BIBdecl

\bibitem[Sharif et~al.(2019)Sharif, Lucas, Bauer, Reiter, and
  Shintre]{2019Optimization}
\BIBentryALTinterwordspacing
M.~Sharif, K.~Lucas, L.~Bauer, M.~K. Reiter, and S.~Shintre,
  ``Optimization-guided binary diversification to mislead neural networks for
  malware detection,'' 2019. [Online]. Available:
  \url{http://arxiv.org/abs/1912.09064}
\BIBentrySTDinterwordspacing

\bibitem[Demetrio et~al.(2021{\natexlab{a}})Demetrio, Biggio, Lagorio, Roli,
  and Armando]{2020Fuctionality}
L.~Demetrio, B.~Biggio, G.~Lagorio, F.~Roli, and A.~Armando,
  ``Functionality-preserving black-box optimization of adversarial windows
  malware,'' \emph{IEEE Transactions on Information Forensics and Security},
  vol.~16, pp. 3469--3478, 2021.

\bibitem[Li et~al.(2015)Li, Ge, and Dai]{2016Detecting}
W.~Li, J.~Ge, and G.~Dai, ``Detecting malware for android platform: An
  svm-based approach,'' in \emph{2015 IEEE 2nd International Conference on
  Cyber Security and Cloud Computing}, 2015, pp. 464--469.

\bibitem[Vinod et~al.(2012)Vinod, Laxmi, Gaur, and Chauhan]{2012MOMENTUM}
P.~Vinod, V.~Laxmi, M.~S. Gaur, and G.~Chauhan, ``Momentum: Metamorphic malware
  exploration techniques using msa signatures,'' in \emph{2012 International
  Conference on Innovations in Information Technology (IIT)}, 2012, pp.
  232--237.

\bibitem[Sharif et~al.(2016)Sharif, Bhagavatula, Bauer, and
  Reiter]{2016Accessorize}
\BIBentryALTinterwordspacing
M.~Sharif, S.~Bhagavatula, L.~Bauer, and M.~K. Reiter, ``Accessorize to a
  crime: Real and stealthy attacks on state-of-the-art face recognition,'' in
  \emph{Proceedings of the 2016 {ACM} {SIGSAC} Conference on Computer and
  Communications Security, Vienna, Austria, October 24-28, 2016}, E.~R. Weippl,
  S.~Katzenbeisser, C.~Kruegel, A.~C. Myers, and S.~Halevi, Eds.\hskip 1em plus
  0.5em minus 0.4em\relax {ACM}, 2016, pp. 1528--1540. [Online]. Available:
  \url{https://doi.org/10.1145/2976749.2978392}
\BIBentrySTDinterwordspacing

\bibitem[{Qin} et~al.(2019){Qin}, {Carlini}, {Goodfellow}, {Cottrell}, and
  {Raffel}]{2019Imperceptible}
Y.~{Qin}, N.~{Carlini}, I.~{Goodfellow}, G.~{Cottrell}, and C.~{Raffel},
  ``{Imperceptible, Robust, and Targeted Adversarial Examples for Automatic
  Speech Recognition},'' p. arXiv:1903.10346, Mar. 2019.

\bibitem[Saxe and Berlin(2015)]{2015Deep}
\BIBentryALTinterwordspacing
J.~Saxe and K.~Berlin, ``Deep neural network based malware detection using two
  dimensional binary program features,'' pp. 11--20, 2015. [Online]. Available:
  \url{https://doi.org/10.1109/MALWARE.2015.7413680}
\BIBentrySTDinterwordspacing

\bibitem[Raff et~al.(2018)Raff, Barker, Sylvester, Brandon, Catanzaro, and
  Nicholas]{article}
\BIBentryALTinterwordspacing
E.~Raff, J.~Barker, J.~Sylvester, R.~Brandon, B.~Catanzaro, and C.~K. Nicholas,
  ``Malware detection by eating a whole {EXE},'' vol. {WS-18}, pp. 268--276,
  2018. [Online]. Available:
  \url{https://aaai.org/ocs/index.php/WS/AAAIW18/paper/view/16422}
\BIBentrySTDinterwordspacing

\bibitem[Anderson et~al.(2018)Anderson, Kharkar, Filar, Evans, and
  Roth]{DBLP:journals/corr/abs-1801-08917}
\BIBentryALTinterwordspacing
H.~S. Anderson, A.~Kharkar, B.~Filar, D.~Evans, and P.~Roth, ``Learning to
  evade static {PE} machine learning malware models via reinforcement
  learning,'' 2018. [Online]. Available: \url{http://arxiv.org/abs/1801.08917}
\BIBentrySTDinterwordspacing

\bibitem[Song et~al.(2020)Song, Li, Afroz, Garg, Kuznetsov, and
  Yin]{DBLP:journals/corr/abs-2003-03100}
\BIBentryALTinterwordspacing
W.~Song, X.~Li, S.~Afroz, D.~Garg, D.~Kuznetsov, and H.~Yin, ``Automatic
  generation of adversarial examples for interpreting malware classifiers,''
  2020. [Online]. Available: \url{https://arxiv.org/abs/2003.03100}
\BIBentrySTDinterwordspacing

\bibitem[Park et~al.(2019)Park, Khan, and
  Yener]{DBLP:journals/corr/abs-1904-04802}
\BIBentryALTinterwordspacing
D.~Park, H.~Khan, and B.~Yener, ``Short paper: Creating adversarial malware
  examples using code insertion,'' 2019. [Online]. Available:
  \url{http://arxiv.org/abs/1904.04802}
\BIBentrySTDinterwordspacing

\bibitem[Arp et~al.(2022)Arp, Quiring, Pendlebury, Warnecke, Pierazzi,
  Wressnegger, Cavallaro, and Rieck]{DBLP:journals/corr/abs-2010-09470}
\BIBentryALTinterwordspacing
D.~Arp, E.~Quiring, F.~Pendlebury, A.~Warnecke, F.~Pierazzi, C.~Wressnegger,
  L.~Cavallaro, and K.~Rieck, ``Dos and don{\textquoteright}ts of machine
  learning in computer security,'' in \emph{31st USENIX Security Symposium
  (USENIX Security 22)}.\hskip 1em plus 0.5em minus 0.4em\relax Boston, MA:
  USENIX Association, Aug. 2022. [Online]. Available:
  \url{https://www.usenix.org/conference/usenixsecurity22/presentation/arp}
\BIBentrySTDinterwordspacing

\bibitem[Nataraj et~al.(2011)Nataraj, Karthikeyan, Jacob, and
  Manjunath]{10.1145/2016904.2016908}
\BIBentryALTinterwordspacing
L.~Nataraj, S.~Karthikeyan, G.~Jacob, and B.~S. Manjunath, ``Malware images:
  Visualization and automatic classification,'' in \emph{Proceedings of the 8th
  International Symposium on Visualization for Cyber Security}, ser. VizSec
  '11.\hskip 1em plus 0.5em minus 0.4em\relax New York, NY, USA: Association
  for Computing Machinery, 2011. [Online]. Available:
  \url{https://doi.org/10.1145/2016904.2016908}
\BIBentrySTDinterwordspacing

\bibitem[Coull and Gardner(2019)]{coull2019activation}
S.~E. Coull and C.~Gardner, ``Activation analysis of a byte-based deep neural
  network for malware classification,'' in \emph{2019 IEEE Security and Privacy
  Workshops (SPW)}, 2019, pp. 21--27.

\bibitem[Johns(2017)]{articletRepresentation}
J.~Johns, ``Representation learning for malware classification,''
  \url{https://www.fireeye.com/content/dam/fireeye-www/blog/pdfs/malware-classification-slides.pdf},
  2017, accessed: 2020-07-26.

\bibitem[Chen et~al.(2019)Chen, Cornelius, Martin, and Chau]{2018Robust}
S.-T. Chen, C.~Cornelius, J.~Martin, and D.~H.~P. Chau, ``Shapeshifter: Robust
  physical adversarial attack on faster r-cnn object detector,'' in
  \emph{Machine Learning and Knowledge Discovery in Databases}, M.~Berlingerio,
  F.~Bonchi, T.~G{\"a}rtner, N.~Hurley, and G.~Ifrim, Eds.\hskip 1em plus 0.5em
  minus 0.4em\relax Cham: Springer International Publishing, 2019, pp. 52--68.

\bibitem[Jia and Liang(2017)]{DBLP:journals/corr/JiaL17}
\BIBentryALTinterwordspacing
R.~Jia and P.~Liang, ``Adversarial examples for evaluating reading
  comprehension systems,'' 2017. [Online]. Available:
  \url{http://arxiv.org/abs/1707.07328}
\BIBentrySTDinterwordspacing

\bibitem[Kreuk et~al.(2018)Kreuk, Barak, Aviv{-}Reuven, Baruch, Pinkas, and
  Keshet]{DBLP:journals/corr/abs-1802-04528}
\BIBentryALTinterwordspacing
F.~Kreuk, A.~Barak, S.~Aviv{-}Reuven, M.~Baruch, B.~Pinkas, and J.~Keshet,
  ``Adversarial examples on discrete sequences for beating whole-binary malware
  detection,'' 2018. [Online]. Available: \url{http://arxiv.org/abs/1802.04528}
\BIBentrySTDinterwordspacing

\bibitem[Suciu et~al.(2019)Suciu, Coull, and Johns]{2019Exploring}
\BIBentryALTinterwordspacing
O.~Suciu, S.~E. Coull, and J.~Johns, ``Exploring adversarial examples in
  malware detection,'' in \emph{2019 {IEEE} Security and Privacy Workshops,
  {SP} Workshops 2019, San Francisco, CA, USA, May 19-23, 2019}.\hskip 1em plus
  0.5em minus 0.4em\relax {IEEE}, 2019, pp. 8--14. [Online]. Available:
  \url{https://doi.org/10.1109/SPW.2019.00015}
\BIBentrySTDinterwordspacing

\bibitem[Pappas et~al.(2012)Pappas, Polychronakis, and Keromytis]{6234439}
V.~Pappas, M.~Polychronakis, and A.~D. Keromytis, ``Smashing the gadgets:
  Hindering return-oriented programming using in-place code randomization,'' in
  \emph{2012 IEEE Symposium on Security and Privacy}, 2012, pp. 601--615.

\bibitem[Ribeiro et~al.(2016)Ribeiro, Singh, and Guestrin]{2016lime}
\BIBentryALTinterwordspacing
M.~T. Ribeiro, S.~Singh, and C.~Guestrin, ``"why should i trust you?":
  Explaining the predictions of any classifier,'' in \emph{Proceedings of the
  22nd ACM SIGKDD International Conference on Knowledge Discovery and Data
  Mining}, ser. KDD '16.\hskip 1em plus 0.5em minus 0.4em\relax New York, NY,
  USA: Association for Computing Machinery, 2016, p. 1135–1144. [Online].
  Available: \url{https://doi.org/10.1145/2939672.2939778}
\BIBentrySTDinterwordspacing

\bibitem[Camburu(2020)]{unknownExplaining}
\BIBentryALTinterwordspacing
O.~Camburu, ``Explaining deep neural networks,'' 2020. [Online]. Available:
  \url{https://arxiv.org/abs/2010.01496}
\BIBentrySTDinterwordspacing

\bibitem[Lundberg and Lee(2017)]{2017A}
\BIBentryALTinterwordspacing
S.~Lundberg and S.~Lee, ``A unified approach to interpreting model
  predictions,'' vol. abs/1705.07874, 2017. [Online]. Available:
  \url{http://arxiv.org/abs/1705.07874}
\BIBentrySTDinterwordspacing

\bibitem[Demetrio et~al.(2019)Demetrio, Biggio, Lagorio, Roli, and
  Armando]{DBLP:journals/corr/abs-1901-03583}
\BIBentryALTinterwordspacing
L.~Demetrio, B.~Biggio, G.~Lagorio, F.~Roli, and A.~Armando, ``Explaining
  vulnerabilities of deep learning to adversarial malware binaries,'' 2019.
  [Online]. Available: \url{http://arxiv.org/abs/1901.03583}
\BIBentrySTDinterwordspacing

\bibitem[Rosenberg et~al.(2020)Rosenberg, Meir, Berrebi, Gordon, Sicard, and
  David]{DBLP:journals/corr/abs-2009-13243}
\BIBentryALTinterwordspacing
I.~Rosenberg, S.~Meir, J.~Berrebi, I.~Gordon, G.~Sicard, and E.~David,
  ``Generating end-to-end adversarial examples for malware classifiers using
  explainability,'' \emph{CoRR}, vol. abs/2009.13243, 2020. [Online].
  Available: \url{https://arxiv.org/abs/2009.13243}
\BIBentrySTDinterwordspacing

\bibitem[Brendel et~al.(2018)Brendel, Rauber, and
  Bethge]{brendel2018decisionbased}
W.~Brendel, J.~Rauber, and M.~Bethge, ``Decision-based adversarial attacks:
  Reliable attacks against black-box machine learning models,'' 2018.

\bibitem[Carlini and Wagner(2017)]{2016Towards}
N.~Carlini and D.~A. Wagner, ``Towards evaluating the robustness of neural
  networks,'' in \emph{2017 IEEE Symposium on Security and Privacy (SP)}, 2017,
  pp. 39--57.

\bibitem[Roweis and K.Saul(2000)]{Roweis00nonlineardimensionality}
S.~T. Roweis and L.~K.Saul, ``Nonlinear dimensionality reduction by locally
  linear embedding,'' \emph{SCIENCE}, vol. 290, pp. 2323--2326, 2000.

\bibitem[{everybodywiki}(2020)]{Ablation2}
{everybodywiki}, ``Ablation analysis,''
  \url{https://en.everybodywiki.com/Ablative\_analysis}, 2020, accessed:
  2021-11-22.

\bibitem[Ghosh et~al.(2019)Ghosh, Das, Das, and
  Maulik]{DBLP:journals/corr/abs-1907-06119}
\BIBentryALTinterwordspacing
S.~Ghosh, N.~Das, I.~Das, and U.~Maulik, ``Understanding deep learning
  techniques for image segmentation,'' 2019. [Online]. Available:
  \url{http://arxiv.org/abs/1907.06119}
\BIBentrySTDinterwordspacing

\bibitem[{Capstone}(2021)]{cap}
{Capstone}, ``The ultimate disassembly framework - capstone- the ultimate
  disassembler,'' \url{https://www.capstone-engine.org}, 2021, accessed:
  2021-03-31.

\bibitem[Fleshman et~al.(2018)Fleshman, Raff, Zak, McLean, and
  Nicholas]{DBLP:journals/corr/abs-1806-04773}
W.~Fleshman, E.~Raff, R.~Zak, M.~McLean, and C.~Nicholas, ``Static malware
  detection amp; subterfuge: Quantifying the robustness of machine learning and
  current anti-virus,'' in \emph{2018 13th International Conference on
  Malicious and Unwanted Software (MALWARE)}, 2018, pp. 1--10.

\bibitem[Koo and Polychronakis(2016)]{2016Juggling}
\BIBentryALTinterwordspacing
H.~Koo and M.~Polychronakis, ``Juggling the gadgets: Binary-level code
  randomization using instruction displacement,'' in \emph{Proceedings of the
  11th ACM on Asia Conference on Computer and Communications Security}, ser.
  ASIA CCS '16.\hskip 1em plus 0.5em minus 0.4em\relax New York, NY, USA:
  Association for Computing Machinery, 2016, p. 23–34. [Online]. Available:
  \url{https://doi.org/10.1145/2897845.2897863}
\BIBentrySTDinterwordspacing

\bibitem[Ronen et~al.(2018)Ronen, Radu, Feuerstein, Yom-Tov, and
  Ahmadi]{ronen2018microsoft}
R.~Ronen, M.~Radu, C.~Feuerstein, E.~Yom-Tov, and M.~Ahmadi, ``Microsoft
  malware classification challenge,'' 2018.

\bibitem[Anderson and Roth(2018)]{2018arXiv180404637A}
\BIBentryALTinterwordspacing
H.~S. Anderson and P.~Roth, ``{EMBER:} an open dataset for training static {PE}
  malware machine learning models,'' 2018. [Online]. Available:
  \url{http://arxiv.org/abs/1804.04637}
\BIBentrySTDinterwordspacing

\bibitem[Krčál et~al.(2018)Krčál, Švec, Bálek, and Jašek]{krl2018deep}
\BIBentryALTinterwordspacing
M.~Krčál, O.~Švec, M.~Bálek, and O.~Jašek, ``Deep convolutional malware
  classifiers can learn from raw executables and labels only,'' 2018. [Online].
  Available: \url{https://openreview.net/forum?id=HkHrmM1PM}
\BIBentrySTDinterwordspacing

\bibitem[Vigna and Balzarotti(2018)]{208119}
\BIBentryALTinterwordspacing
G.~Vigna and D.~Balzarotti, ``When malware is {Packin{\textquoteright}} heat,''
  in \emph{Enigma 2018 (Enigma 2018)}.\hskip 1em plus 0.5em minus 0.4em\relax
  Santa Clara, CA: USENIX Association, Jan. 2018. [Online]. Available:
  \url{https://www.usenix.org/node/208120}
\BIBentrySTDinterwordspacing

\bibitem[Noever and Miller.Noever(2021)]{noever2021virusmnist}
D.~Noever and S.~E. Miller.Noever, ``Virus-mnist:a benchark malware dataset,''
  2021.

\bibitem[Yang et~al.(2021)Yang, Ciptadi, Laziuk, Ahmadzadeh, and Wang]{9474321}
L.~Yang, A.~Ciptadi, I.~Laziuk, A.~Ahmadzadeh, and G.~Wang, ``Bodmas: An open
  dataset for learning based temporal analysis of pe malware,'' in \emph{2021
  IEEE Security and Privacy Workshops (SPW)}, 2021, pp. 78--84.

\bibitem[Kolosnjaji et~al.(2018)Kolosnjaji, Demontis, Biggio, Maiorca,
  Giacinto, Eckert, and Roli]{8553214}
B.~Kolosnjaji, A.~Demontis, B.~Biggio, D.~Maiorca, G.~Giacinto, C.~Eckert, and
  F.~Roli, ``Adversarial malware binaries: Evading deep learning for malware
  detection in executables,'' in \emph{2018 26th European Signal Processing
  Conference (EUSIPCO)}, 2018, pp. 533--537.

\bibitem[{Microsoft}(2021)]{c22}
{Microsoft}, ``Pe format,''
  \url{https://docs.microsoft.com/en-us/windows/win32/debug/pe-format}, 2021,
  accessed: 2021-03-31.

\bibitem[Lucas et~al.(2021)Lucas, Sharif, Bauer, Reiter, and
  Shintre]{10.1145/3433210.3453086}
\BIBentryALTinterwordspacing
K.~Lucas, M.~Sharif, L.~Bauer, M.~K. Reiter, and S.~Shintre, ``Malware
  makeover: Breaking ml-based static analysis by modifying executable bytes,''
  in \emph{Proceedings of the 2021 ACM Asia Conference on Computer and
  Communications Security}, ser. ASIA CCS '21.\hskip 1em plus 0.5em minus
  0.4em\relax New York, NY, USA: Association for Computing Machinery, 2021, p.
  744–758. [Online]. Available: \url{https://doi.org/10.1145/3433210.3453086}
\BIBentrySTDinterwordspacing

\bibitem[{Oleh Yuschuk}(2014)]{ollydbg}
{Oleh Yuschuk}, ``Ollydbg,'' \url{https://www.ollydbg.de}, 2014, accessed:
  2022-2-10.

\bibitem[{Claudio Guarnieri, Allessandro Tanasi, Jurriaan Bremer, and Mark
  Schloesser}(2019)]{cuckoo}
{Claudio Guarnieri, Allessandro Tanasi, Jurriaan Bremer, and Mark Schloesser},
  ``Cuckoo sandbox,'' \url{https://www.cuckoosandbox.org}, 2019, accessed:
  2022-2-10.

\bibitem[Lowd and Meek(2005)]{10.11451081870.1081950}
\BIBentryALTinterwordspacing
D.~Lowd and C.~Meek, ``Adversarial learning,'' in \emph{Proceedings of the
  Eleventh ACM SIGKDD International Conference on Knowledge Discovery in Data
  Mining}, ser. KDD '05.\hskip 1em plus 0.5em minus 0.4em\relax New York, NY,
  USA: Association for Computing Machinery, 2005, p. 641–647. [Online].
  Available: \url{https://doi.org/10.1145/1081870.1081950}
\BIBentrySTDinterwordspacing

\bibitem[Szegedy et~al.(2014)Szegedy, Zaremba, Sutskever, Bruna, Erhan,
  Goodfellow, and Fergus]{DBLPjournalscorrSzegedyZSBEGF13}
\BIBentryALTinterwordspacing
C.~Szegedy, W.~Zaremba, I.~Sutskever, J.~Bruna, D.~Erhan, I.~J. Goodfellow, and
  R.~Fergus, ``Intriguing properties of neural networks,'' in \emph{2nd
  International Conference on Learning Representations, {ICLR} 2014, Banff, AB,
  Canada, April 14-16, 2014, Conference Track Proceedings}, Y.~Bengio and
  Y.~LeCun, Eds., 2014. [Online]. Available:
  \url{http://arxiv.org/abs/1312.6199}
\BIBentrySTDinterwordspacing

\bibitem[Papernot et~al.(2018)Papernot, McDaniel, Sinha, and Wellman]{8406613}
N.~Papernot, P.~McDaniel, A.~Sinha, and M.~P. Wellman, ``Sok: Security and
  privacy in machine learning,'' in \emph{2018 IEEE European Symposium on
  Security and Privacy (EuroS P)}, 2018, pp. 399--414.

\bibitem[Guo et~al.(2018)Guo, Mu, Xu, Su, Wang, and Xing]{Guo2018LEMNAED}
\BIBentryALTinterwordspacing
W.~Guo, D.~Mu, J.~Xu, P.~Su, G.~Wang, and X.~Xing, ``Lemna: Explaining deep
  learning based security applications,'' p. 364–379, 2018. [Online].
  Available: \url{https://doi.org/10.1145/3243734.3243792}
\BIBentrySTDinterwordspacing

\bibitem[Demetrio et~al.(2021{\natexlab{b}})Demetrio, Coull, Biggio, Lagorio,
  Armando, and Roli]{10.11453473039}
\BIBentryALTinterwordspacing
L.~Demetrio, S.~E. Coull, B.~Biggio, G.~Lagorio, A.~Armando, and F.~Roli,
  ``Adversarial exemples: A survey and experimental evaluation of practical
  attacks on machine learning for windows malware detection,'' \emph{ACM Trans.
  Priv. Secur.}, vol.~24, no.~4, sep 2021. [Online]. Available:
  \url{https://doi.org/10.1145/3473039}
\BIBentrySTDinterwordspacing

\bibitem[Pope et~al.(2021)Pope, Zhu, Abdelkader, Goldblum, and
  Goldstein]{pope2021the}
\BIBentryALTinterwordspacing
P.~Pope, C.~Zhu, A.~Abdelkader, M.~Goldblum, and T.~Goldstein, ``The intrinsic
  dimension of images and its impact on learning,'' 2021. [Online]. Available:
  \url{https://openreview.net/forum?id=XJk19XzGq2J}
\BIBentrySTDinterwordspacing

\bibitem[{Johannes Plachy}(2018)]{c23}
{Johannes Plachy}, ``Portable executable file format,''
  \url{https://blog.kowalczyk.info/articles/pefileformat.html}, 2018, accessed:
  2018-7-26.

\end{thebibliography}
\section{Appendix}
\subsection{Windows Portable Executable File Format}

The data we use in this article are all Windows PE files and we take advantage of the format characteristics of the PE files to create adversarial examples. The PE files are derived from the Common Object File Format (COFF), which specifies how Windows executables are stored on the disk. The main file that specifies the PE files is winnt.h, related documents can also be found in \cite{c22}. There are two types of PE files, one is executable (EXE) file and the other is dynamic link library (DLL) file. They are almost the same in terms of file format, the only difference is that a field is used to identify whether the file is an EXE or DLL. Generally, PE files can be roughly divided into different components. They begin with a MS-DOS header, a Stub and a PE file signature. Immediately following is the PE file header and optional header. Beyond that, section headers and section bodies follow. A PE file typically has nine predefined sections named .text,.bss,.rdata,.data,.rsrc,.edata,.idata,.pdata and .debug \cite{c23}. Some binaries do not need all of these sections while others may rename or define the section names according to their own needs. For the alignment reason, the start address of the segment part of PE is often 0x100. 
\begin{figure}[h]
	\centering
	\includegraphics[width=.4\columnwidth]{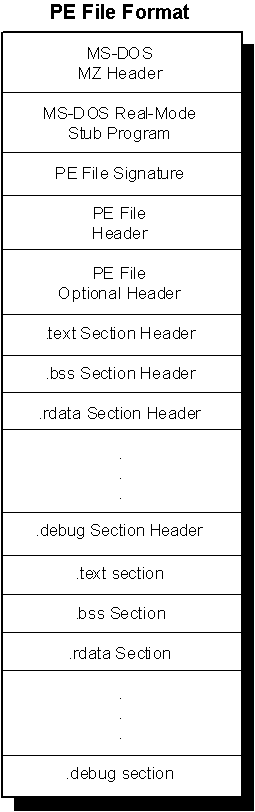}
	\caption{Structure of a typical PE file image.	}
	\label{fig:Structure of a PE file image.}
\end{figure}
\end{document}